\documentclass[manuscript,screen,nonacm,11pt]{acmart}
\AtBeginDocument{%
  }

\usepackage{graphicx}
\usepackage{subcaption} 

\usepackage{multirow} 

\usepackage{algorithm}
\usepackage{algpseudocode}

\setcopyright{acmlicensed}
\copyrightyear{2018}
\acmYear{2018}
\acmDOI{XXXXXXX.XXXXXXX}
\acmConference[Conference acronym 'XX]{Make sure to enter the correct
  conference title from your rights confirmation email}{June 03--05,
  2018}{Woodstock, NY}
\acmISBN{978-1-4503-XXXX-X/2018/06}

\begin{document}

\title{Task-Aware Delegation Cues for LLM Agents}

\author{Xingrui Gu}
\email{xingrui\_gu@berkeley.edu}
\orcid{0009-0009-6415-9268}
\affiliation{%
  \institution{University of California, Berkeley}
  \city{Berkeley}
  \state{CA}
  \country{USA}
}

\renewcommand{\shortauthors}{Xingrui Gu}

\begin{abstract}
LLM agents increasingly present as conversational collaborators, yet human--agent teamwork remains brittle due to information asymmetry: users lack task-specific reliability cues, and agents rarely surface calibrated uncertainty or rationale. We propose a task-aware collaboration signaling layer that turns offline preference evaluations into online, user-facing primitives for delegation. Using Chatbot Arena pairwise comparisons, we induce an interpretable task taxonomy via semantic clustering, then derive (i) Capability Profiles as task-conditioned win-rate maps and (ii) Coordination-Risk Cues as task-conditioned disagreement (tie-rate) priors. These signals drive a closed-loop delegation protocol that supports common-ground verification, adaptive routing (primary vs.\ primary+auditor), explicit rationale disclosure, and privacy-preserving accountability logs. Two predictive probes validate that task typing carries actionable structure: cluster features improve winner prediction accuracy and reduce difficulty prediction error under stratified 5-fold cross-validation. Overall, our framework reframes delegation from an opaque system default into a visible, negotiable, and auditable collaborative decision, providing a principled design space for adaptive human--agent collaboration grounded in mutual awareness and shared accountability.
\end{abstract}

\received[accepted]{26 Feb 2026}

\maketitle

\section{Introduction}

LLM-based agent systems have quickly moved from “tools” to conversational collaborators capable of dialogue, reasoning, and multi-step planning \citep{dhillon2024shaping, yao2022react, schick2023toolformer, park2023generative}. Yet collaborator-like behavior does not guarantee effective collaboration \citep{zhang2021ideal, amershi2019guidelines}. Relative to human teams, human–agent interaction still lacks core collaborative properties—mutual awareness, adaptivity, and shared accountability \citep{mitchell2019model, gu2025causkelnet, gu2024advancing}. Users often cannot assess an agent’s task-specific competence, reliability, or failure modes \citep{mitchell2019model}, while agents rarely surface calibrated uncertainty or decision rationale \citep{xu2025confronting}. The resulting opacity makes interactions brittle and failures hard to diagnose or repair \citep{shah2018algorithmic}. In distributed teamwork, collaborators compensate for limited shared context through explicit signaling to establish common ground and accountability \citep{clark1991grounding, duckert2023collocated}. Analogously, human–agent collaboration is characterized by information asymmetry—users lack actionable reliability cues and agents lack standardized channels to communicate rationale and uncertainty—leading to trust miscalibration and an accountability vacuum when errors occur \citep{lee2004trust, amershi2019guidelines, santoni2018meaningful}.

A central challenge is deciding which agent is most reliable for a given delegation and when to trigger interventions such as clarification or dual verification \citep{huang2025routereval, chen2023frugalgpt, horvitz1999principles, testoni2024asking, dhuliawala2024chain}. Existing methods often depend on coarse global rankings that miss task-specific brittleness—models may excel in one domain yet hallucinate in another—and rarely adapt to intrinsic task ambiguity \citep{chiang2024chatbot, testoni2024asking}.

\begin{figure}[t]
    \centering
    \includegraphics[width=\linewidth]{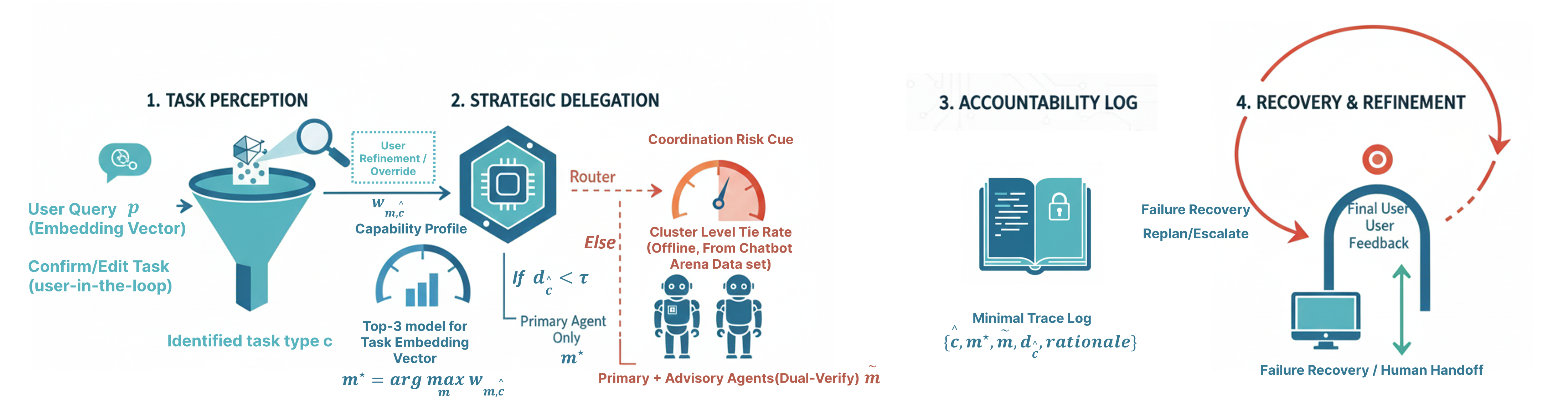}
    \caption{\textbf{The Task-Aware Delegation \& Awareness Loop.} Our framework operationalizes (1) Intent Recognition via semantic clustering, (2) Dynamic Delegation based on capability profiles, (3) Awareness Cues for trust calibration, and (4) Accountability Logging for error recovery.}
    \label{fig:loop}
\end{figure}

We propose a task-aware collaboration signaling layer that turns offline capability assessments into online, user-facing cues. Combining semantic task clustering with large-scale human preference data, we derive task-conditioned Capability Profiles and Uncertainty Cues, which drive a closed-loop delegation protocol (task typing $\rightarrow$ routing $\rightarrow$ explicit rationale $\rightarrow$ accountable logging). This reframes delegation from an opaque default into a visible, negotiable, and auditable collaborative decision, supporting calibrated trust and adaptive human–agent teamwork \citep{amershi2019guidelines, mitchell2019model, shah2018algorithmic}.

\section{Related Work: Agents as Remote Collaborators}
A burgeoning perspective in Human-Computer Interaction (HCI) frames intelligent agents not merely as tools, but as \textit{remote collaborators} \citep{amershi2019guidelines}. This conceptual shift draws on classic CSCW insights into distributed teamwork, where the absence of a shared physical environment necessitates a reliance on explicit signaling to establish \textbf{common ground} \citep{clark1991grounding,olson2000distance}. As defined by Media Richness Theory, text-based interaction---despite using natural language---often functions as a relatively ``lean'' channel compared to face-to-face collaboration \citep{daft1986organizational}. In such environments, the agent's internal reasoning, capability boundaries, and uncertainty are often obscured, leading to systemic information asymmetry and making appropriate reliance difficult \citep{amershi2019guidelines,mitchell2019model,lee2004trust, gu2026laplacian}. Previous studies have shown that reduced shared context and limited cues can increase grounding/coordination costs in remote collaboration \citep{fussell2000coordination, olson2000distance, gu2024advancing, gu2025causkelnet}, while miscalibrated trust can result in misuse/disuse when failures occur \citep{parasuraman1997humans, lee2004trust, guuncertainty, gu2024mimicking}. This theoretical backdrop suggests that for LLM agents to become genuine partners, they must move beyond task execution toward proactive signaling of their task-specific reliability and context.

\section{Preference-Derived Collaboration Signals}
\label{sec:signals}
We derive task-conditioned collaboration signals from human preference comparisons, capturing
(i) \emph{capability} (who tends to win on a task type) and (ii) \emph{coordination risk}
(when disagreement is high and additional grounding/auditing is warranted).

\subsection{Task typing}
\label{sec:task_typing}

Given prompts $\mathcal{P}=\{p_i\}_{i=1}^N$, we obtain semantic embeddings
\begin{equation}
e_i = f_{\mathrm{enc}}(p_i) \in \mathbb{R}^d,
\end{equation}
apply a standard dimensionality reduction operator $g(\cdot)$ (e.g., UMAP), and perform K-means clustering
to obtain a task type $c_i \in \{1,\dots,K\}$ for each prompt:
\begin{equation}
x_i = g(e_i),\qquad c_i = \mathrm{KMeans}(x_i), \qquad K=30.
\end{equation}
Equivalently, K-means solves
\begin{equation}
\arg\min_{\{C_j\}_{j=1}^K}\ \sum_{j=1}^K \sum_{x_i\in C_j}\|x_i-\mu_j\|_2^2,
\end{equation}
where $\mu_j$ is the centroid of cluster $C_j$. We optionally assign human-readable labels to clusters via representative keywords for presentation. Figure~\ref{fig:task_space} visualizes the induced task-typing structure in a low-dimensional projection.

\subsection{Capability profile}
\label{sec:capability}

Each prompt $p_i$ is associated with a human comparison record $(p_i, m_A, m_B, y_i)$ where
$y_i \in \{m_A, m_B, \mathrm{tie}\}$.
For each agent/model $m$ and task cluster $c$, we define the empirical win rate
\begin{equation}
w_{m,c} \;=\; \frac{1}{|\mathcal{D}_{m,c}|}\sum_{(i,y_i)\in \mathcal{D}_{m,c}} \mathbb{I}[y_i = m],
\label{eq:winrate}
\end{equation}
where $\mathcal{D}_{m,c}$ denotes comparisons involving $m$ on prompts with $c_i=c$. We treat $w_{m,c}$ as a task-conditioned capability signal. Figure~\ref{fig:capability_map} shows that winner distributions differ substantially across task types, supporting task-conditioned capability profiling.

\subsection{Coordination risk cue}
\label{sec:coord_risk}

We quantify task-conditioned uncertainty via the tie rate within each cluster:
\begin{equation}
d_c \;=\; \frac{1}{|\mathcal{D}_c|}\sum_{(i,y_i)\in \mathcal{D}_{c}} \mathbb{I}[y_i = \mathrm{tie}],
\label{eq:tierate}
\end{equation}
where $\mathcal{D}_c$ collects all comparisons with $c_i=c$.
We interpret $d_c$ as a \emph{coordination-risk cue} (model disagreement / ambiguity), rather than a ground-truth hardness measure. Figure~\ref{fig:coord_risk} reports the task-type-level uncertainty proxy used as a coordination-risk cue.

\section{Experiments: Validating Task-Conditioned Collaboration Signals}
\label{sec:experiments}

Section~\ref{sec:signals} introduces two task-conditioned signals from preference data: a \emph{capability profile} (task-specific win tendency) and a \emph{coordination-risk cue} (high disagreement/uncertainty). We evaluate their behavioral validity using two probes on Chatbot-Arena-style pairwise comparisons: (A) \textbf{winner prediction} and (B) \textbf{difficulty prediction}. Task~A tests whether task typing explains systematic, task-dependent performance differences; Task~B tests whether task type and disagreement jointly predict perceived difficulty, justifying disagreement-based risk cues.

\paragraph{\textbf{Dataset and evaluation protocol.}}
We use the Chatbot Arena dataset \citep{chiang2024chatbot}, which consists of single-turn prompts paired with human preference comparisons between two LLMs, denoted by $(p_i, m_A, m_B, y_i)$, where $y_i \in \{m_A, m_B, \mathrm{tie}, \mathrm{invalid}\}$. Each record also includes prompt metadata and embedding features. We report stratified 5-fold cross-validation (CV) performance and compare regularization choices across models.

\paragraph{\textbf{Task A: Model winner classification.}} Task~A predicts the outcome of a pairwise comparison (A wins / B wins / tie / invalid). We use multinomial logistic regression for interpretability. Feature engineering follows three groups: (i) one-hot model identities for both contestants, (ii) one-hot task cluster $c_i$ from Section~\ref{sec:task_typing}, and (iii) response-embedding difference $e_A - e_B$.

\paragraph{\textbf{Task B: Prompt difficulty prediction.}} Task~B predicts a prompt difficulty score (1--10). We use Ridge regression and evaluate with MSE. Features include (i) task cluster identity (capturing task-type-level difficulty trends), (ii) winner-combination indicators (capturing disagreement patterns), and (iii) prompt length.

\paragraph{\textbf{Results.}}

Table~\ref{tab:cv_summary} shows that Ridge regularization performs best for both tasks under 5-fold CV, which also isolates the contribution of task typing: removing cluster features reduces Task~A accuracy and increases Task~B MSE, supporting the predictive value of task-conditioned
signals for both capability and coordination risk.

\section{Delegation Protocol}
We operationalize the task-conditional capability profiles $w_{m,c}$ (Eq.~\ref{eq:winrate}) and collaborative risk cues $d_c$ (Eq.~\ref{eq:tierate}) into an online delegation process, formalized in Algorithm~\ref{alg:delegation}. Here, the cluster-level tie-rate $d_c$, defined in Section~\ref{sec:coord_risk}, serves as a proxy for uncertainty and disagreement to characterize task-conditional coordination risks. Given a user request $p$, the system first predicts the task category $\hat c$, allowing for user overrides to establish \textit{common ground}. Subsequently, the system selects the primary collaborator $m^\star(\hat c) = \arg\max_m w_{m,\hat c}$. If the risk signal exceeds a threshold ($d_{\hat c} > \tau$), a high-assurance mode is triggered (e.g., seeking clarification, auditing, or grounding), potentially assigning an auxiliary auditor $\tilde m(\hat c)$. Before execution, the system explicitly surfaces delegation rationales based on $(w_{m,\hat c}, d_{\hat c})$ and maintains a minimized, privacy-preserving \textit{accountability log} to support error recovery and retrospective auditing. In high-disagreement scenarios ($d_{\hat c} > \tau$), the system proactively poses a clarifying question and enables cross-verification by $\tilde m$. Conversely, in low-disagreement tasks ($d_{\hat c} \le \tau$), the system prioritizes efficiency by executing the task directly through $m^\star$ to minimize coordination overhead.

\begin{algorithm}[t]
\caption{Task-aware Delegation with Capability \& Coordination-Risk Cues}
\label{alg:delegation}
\small
\begin{algorithmic}[1]
\Require user request $p$; task typer $\hat c=\mathrm{Type}(p)$; capability $w_{m,c}$ (Eq.~\ref{eq:winrate});
risk cue $d_c$ (tie-rate, Eq.~\ref{eq:tierate}); threshold $\tau$
\State \textbf{Type \& verify:} show $\hat c$ (with rationale/confidence); allow user override $\hat c \leftarrow \mathrm{Edit}(\hat c)$
\State \textbf{Delegate:} $m^\star \leftarrow \arg\max_m w_{m,\hat c}$ \Comment{primary collaborator}
\If{$d_{\hat c} > \tau$}
    \State $\tilde m \leftarrow \arg\max_{m\neq m^\star} w_{m,\hat c}$ \Comment{auditor / backup}
    \State trigger safeguard(s): \{one clarification, auditing, cite sources, stepwise plan\}
\Else
    \State $\tilde m \leftarrow \varnothing$
\EndIf
\State \textbf{Expose awareness cues:} report delegation rationale using $(w_{m,\hat c}, d_{\hat c})$ and active strategy
\State \textbf{Accountability log:} record $\{\hat c, m^\star, \tilde m, d_{\hat c}, \text{safeguards}, \text{repair/handoff}\}$ under privacy constraints
\State \textbf{Execute:} generate output with $m^\star$; optionally audit with $\tilde m$
\end{algorithmic}
\end{algorithm}

\subsection{Risks and Safeguards.}
(1) \textbf{Bias Amplification:} Semantic clustering may \textbf{reify} representational biases. To counteract this, our protocol mandates user-editable task labels and provides interpretable classification rationales to support human-in-the-loop auditing. 
(2) \textbf{Trust Miscalibration:} High performance scores ($w_{m,c}$) can induce user over-reliance. We mitigate this by mandating the disclosure of model limitations and surfacing risk-adjusted strategies whenever $d_{\hat c} > \tau$, fostering a more critical and calibrated trust. 
(3) \textbf{Privacy Inference:} Aggregated task logs could potentially facilitate sensitive user profiling. The protocol enforces \textbf{minimalist data retention} by default, provides granular ``right-to-be-forgotten'' mechanisms, and applies noise to logging frequency in high-sensitivity task clusters.


\newpage
\bibliographystyle{ACM-Reference-Format}
\bibliography{sample-base}


\appendix

\newpage
\section{Implementation Details: Task Typing and Preference Modeling}
\label{app:impl}

This appendix details the end-to-end implementation of task typing (semantic clustering),
preference alignment, and the two predictive probes (Task~A/B) used in Sec.~\ref{sec:experiments}.
Our goal is reproducibility: we specify inputs, preprocessing, hyperparameters, and evaluation.

\subsection{Inputs and data representation}
\label{app:inputs}

We assume a dataset of prompts and pairwise preference comparisons:
\[
\mathcal{P}=\{p_i\}_{i=1}^N,\qquad 
\mathcal{D}=\{(p_i,m_A,m_B,y_i)\},
\]
where $y_i\in\{m_A,m_B,\mathrm{tie},\mathrm{invalid}\}$.
Each prompt is assigned a task type $c_i\in\{1,\dots,K\}$ via semantic clustering (Sec.~\ref{app:clustering}),
and comparisons are aligned to clusters for computing task-conditioned statistics (Sec.~\ref{app:alignment}).

\subsection{Semantic embedding and task clustering}
\label{app:clustering}

\paragraph{Prompt embeddings.}
We embed each prompt using a pretrained sentence encoder $f_{\mathrm{enc}}$ (Sentence-BERT):
\begin{equation}
e_i = f_{\mathrm{enc}}(p_i)\in\mathbb{R}^d.
\end{equation}

\paragraph{Dimensionality reduction.}
To improve clusterability and reduce noise, we apply a standard manifold learning operator $g(\cdot)$
(e.g., UMAP) to obtain $x_i=g(e_i)\in\mathbb{R}^{d'}$ with $d'\ll d$.
For completeness, UMAP optimizes a cross-entropy objective that preserves neighborhood structure; we
refer to the UMAP paper for the full derivation.

\paragraph{Clustering.}
We run $K$-means on $\{x_i\}$ with $K=30$ to obtain clusters $\{C_j\}_{j=1}^K$:
\begin{equation}
\arg\min_{\{C_j\}_{j=1}^K}\ \sum_{j=1}^K\sum_{x_i\in C_j}\|x_i-\mu_j\|_2^2,
\end{equation}
where $\mu_j$ is the centroid of cluster $C_j$.
To reduce fragmentation, clusters with size below a threshold $\delta$ are reassigned to the nearest
large cluster by centroid distance:
\begin{equation}
C_{\mathrm{target}}
=\arg\min_{j\in\mathcal{L}}
\|\mu_{\mathrm{small}}-\mu_j\|_2,
\end{equation}
where $\mathcal{L}$ indexes clusters with size $\ge\delta$.

\paragraph{Cluster presentation.}
For interpretability in the UI/protocol layer, we optionally attach a short label to each cluster
via representative keywords (e.g., top-frequency tokens after standard text preprocessing).

\subsection{Preference alignment and signal computation}
\label{app:alignment}

Each comparison $(p_i,m_A,m_B,y_i)\in\mathcal{D}$ is aligned to the cluster of its prompt $c_i$.
We compute two task-conditioned signals used in Sec.~\ref{sec:signals}:

\paragraph{Capability profile (win-rate).}
For each model $m$ and cluster $c$, we compute
\begin{equation}
w_{m,c}=\frac{1}{|\mathcal{D}_{m,c}|}\sum_{(i,y_i)\in\mathcal{D}_{m,c}}\mathbb{I}[y_i=m],
\end{equation}
where $\mathcal{D}_{m,c}$ denotes comparisons involving $m$ whose prompt has $c_i=c$.

\paragraph{Coordination-risk cue (tie-rate).}
We quantify task-conditioned uncertainty/disagreement via the cluster tie-rate:
\begin{equation}
d_c=\frac{1}{|\mathcal{D}_c|}\sum_{(i,y_i)\in\mathcal{D}_c}\mathbb{I}[y_i=\mathrm{tie}],
\end{equation}
where $\mathcal{D}_c$ contains all comparisons with $c_i=c$.
In the main paper, $d_c$ is treated as a coordination-risk proxy rather than a ground-truth hardness label.

\subsection{Pipeline pseudocode}
\label{app:pseudocode}

\begin{algorithm}[t]
\caption{End-to-end Pipeline: Task Typing $\rightarrow$ Preference Signals $\rightarrow$ Validation Probes}
\label{alg:pipeline}
\small
\begin{algorithmic}[1]
\Require Prompts $\mathcal{P}=\{p_i\}_{i=1}^N$; pairwise comparisons $\mathcal{R}=\{(p_i,m_A,m_B,y_i)\}$ with $y_i\in\{m_A,m_B,\mathrm{tie},\mathrm{invalid}\}$; encoder $f_{\mathrm{enc}}$; reducer $g$ (e.g., UMAP); $K$ clusters; CV folds $F$
\Ensure Task labels $\{c_i\}$; capability profiles $\{w_{m,c}\}$; coordination-risk cues $\{d_c\}$; CV metrics for Task A/B

\Statex \textbf{Step 1: Task typing (Sec.~\ref{sec:task_typing})}
\For{$i=1$ to $N$}
    \State $e_i \leftarrow f_{\mathrm{enc}}(p_i)$
    \State $x_i \leftarrow g(e_i)$
\EndFor
\State $\{c_i\}_{i=1}^N \leftarrow \mathrm{KMeans}(\{x_i\}, K)$
\State Optionally: assign human-readable labels to clusters via representative keywords

\Statex \textbf{Step 2: Preference-derived signals (Sec.~\ref{sec:capability}--\ref{sec:coord_risk})}
\For{each cluster $c \in \{1,\dots,K\}$}
    \State $\mathcal{D}_c \leftarrow \{(i,y_i)\in\mathcal{R}: c_i=c\}$
    \State $d_c \leftarrow \frac{1}{|\mathcal{D}_c|}\sum_{(i,y_i)\in\mathcal{D}_c}\mathbb{I}[y_i=\mathrm{tie}]$ \Comment{Eq.~\ref{eq:tierate}}
\EndFor
\For{each model/agent $m$ and cluster $c$}
    \State $\mathcal{D}_{m,c} \leftarrow \{(i,y_i)\in\mathcal{R}: c_i=c \ \wedge\ (m=m_A \vee m=m_B)\}$
    \State $w_{m,c} \leftarrow \frac{1}{|\mathcal{D}_{m,c}|}\sum_{(i,y_i)\in\mathcal{D}_{m,c}}\mathbb{I}[y_i=m]$ \Comment{Eq.~\ref{eq:winrate}}
\EndFor

\Statex \textbf{Step 3: Validation probes (Sec.~\ref{sec:experiments})}
\State Partition data into stratified $F$ folds by outcome label for cross-validation
\For{$f=1$ to $F$}
    \State $\mathcal{R}_{\mathrm{train}}, \mathcal{R}_{\mathrm{test}} \leftarrow \mathrm{Split}(\mathcal{R}, f)$

    \Statex \quad \textbf{Task A (winner prediction).}
    \State Construct features $\phi_A(i)$ for each record in $\mathcal{R}_{\mathrm{train}}$:
           one-hot model IDs $(m_A,m_B)$ + one-hot cluster $c_i$ + response-embedding diff $(e_A-e_B)$
    \State Train multinomial logistic regression (optionally with regularization) on $\mathcal{R}_{\mathrm{train}}$
    \State Evaluate on $\mathcal{R}_{\mathrm{test}}$ and record accuracy

    \Statex \quad \textbf{Task B (difficulty prediction).}
    \State Construct features $\phi_B(i)$ for each record in $\mathcal{R}_{\mathrm{train}}$:
           one-hot cluster $c_i$ + winner-combination indicators + prompt length
    \State Train ridge regression (optionally compare with none/lasso) on $\mathcal{R}_{\mathrm{train}}$
    \State Evaluate on $\mathcal{R}_{\mathrm{test}}$ and record MSE
\EndFor
\State Report mean CV metrics across folds; optionally run cluster-feature ablation by removing one-hot $c_i$ and re-evaluating

\end{algorithmic}
\end{algorithm}

\begin{algorithm}[t]
\caption{Task typing and preference signal computation}
\label{alg:pipeline}
\small
\begin{algorithmic}[1]
\Require Prompts $\mathcal{P}$, comparisons $\mathcal{D}$, encoder $f_{\mathrm{enc}}$, reducer $g(\cdot)$, $K$, threshold $\delta$
\Ensure Task types $\{c_i\}$, capability $w_{m,c}$, risk cue $d_c$
\For{$p_i\in\mathcal{P}$}
  \State $e_i\leftarrow f_{\mathrm{enc}}(p_i)$; \quad $x_i\leftarrow g(e_i)$
\EndFor
\State $\{C_j\}_{j=1}^K \leftarrow \mathrm{Cluster\_Method(K-Means)}(\{x_i\},K)$
\State Reassign small clusters ($|C_j|<\delta$) to nearest large cluster by centroid distance
\For{$(p_i,m_A,m_B,y_i)\in\mathcal{D}$}
  \State $c_i\leftarrow \mathrm{cluster}(p_i)$ \Comment{prompt $\to$ task type}
  \State Align vote outcome to $(m_A,c_i)$ and $(m_B,c_i)$
\EndFor
\State Compute $w_{m,c}$ and $d_c$ by aggregated counts over aligned comparisons
\end{algorithmic}
\end{algorithm}

\subsection{Experiment details: predictive probes (Task A/B)}
\label{app:exp_details}

We include two predictive probes to validate that task typing carries signal beyond global model identity.

\paragraph{Task A: winner prediction (classification).}
Task~A is a 4-class classification problem predicting $y_i\in\{\text{A wins},\text{B wins},\text{tie},\text{invalid}\}$.
We use multinomial logistic regression for interpretability.
\emph{Invalid} comparisons can be either (i) retained as a separate class or (ii) excluded from training;
we follow the main-text setting and report results accordingly.

\emph{Features.} We concatenate three feature groups:
(i) model identity one-hots for the two contestants (40 binary features for 20 LLMs);
(ii) task cluster one-hot (30 binary features); and
(iii) response-embedding difference $e_{\mathrm{diff}}=e_A-e_B$ (256 numeric features),
yielding a 326-dimensional vector.

\paragraph{Task B: difficulty prediction (regression).}
Task~B predicts a scalar prompt difficulty score (1--10). We use Ridge regression and evaluate by MSE.
\emph{Features.} We use (i) task cluster one-hot (30),
(ii) winner-combination indicators (5),
and (iii) prompt length (1), for a 36-dimensional vector.

\paragraph{Evaluation protocol.}
We report stratified 5-fold cross-validation. We compare regularization choices (None / Ridge / Lasso)
and include an ablation that removes the cluster one-hot features to quantify the contribution of task typing.

\begin{table}[t]
\centering
\small
\setlength{\tabcolsep}{5pt}
\begin{tabular}{lccccccc}
\toprule
& \multicolumn{3}{c}{5-fold CV (Regularization)} 
& \multicolumn{3}{c}{5-fold CV (Ablation; Ridge)} \\
\cmidrule(lr){2-4}\cmidrule(lr){5-7}
Task & None & Ridge & Lasso & With cluster & Without cluster & $\Delta$ \\
\midrule
A (Acc $\uparrow$) & 0.543 & \textbf{0.548} & 0.545 & \textbf{0.548} & 0.541 & +0.007 \\
B (MSE $\downarrow$) & 2.465 & \textbf{2.463} & 3.664 & \textbf{2.463} & 2.567 & -0.104 \\
\bottomrule
\end{tabular}
\caption{\textbf{5-fold cross-validation results.} Left: regularization sweep for Task A (winner prediction) and Task B (difficulty prediction). Right: effect of task-typing (cluster) features under Ridge (best-performing regularizer).}
\label{tab:cv_summary}
\end{table}



\section{Additional Visualizations}

\begin{figure}[t]
  \centering
  \includegraphics[width=0.9\linewidth]{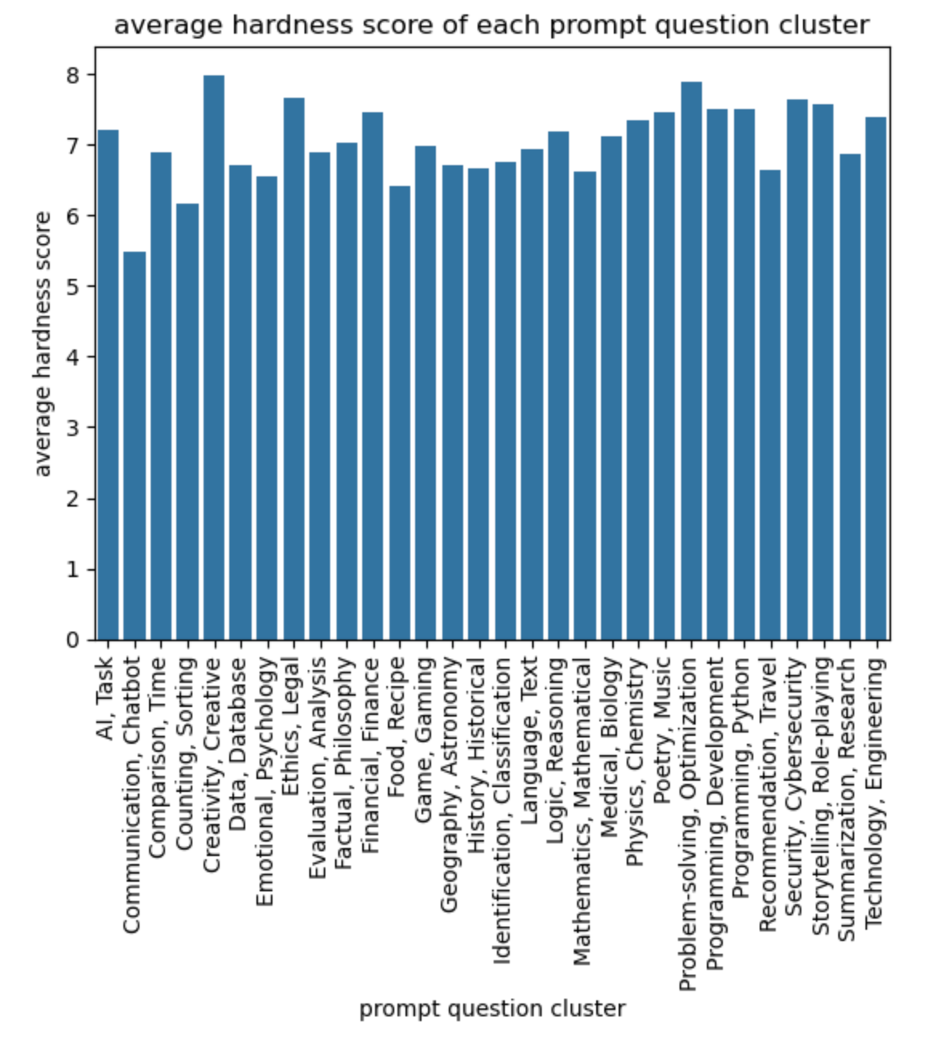}
  \caption{\textbf{Uncertainty / Coordination Risk by Task Type.}
  We report a task-type-level proxy for uncertainty (e.g., tie rate / disagreement-derived hardness) aggregated within
  each cluster. Higher values indicate stronger model disagreement and thus elevated coordination risk, motivating
  safeguards such as clarification and auditing (Sec.~\ref{sec:protocol}).}
  \label{fig:coord_risk}
\end{figure}

\begin{figure}[t]
  \centering
  \includegraphics[width=\linewidth]{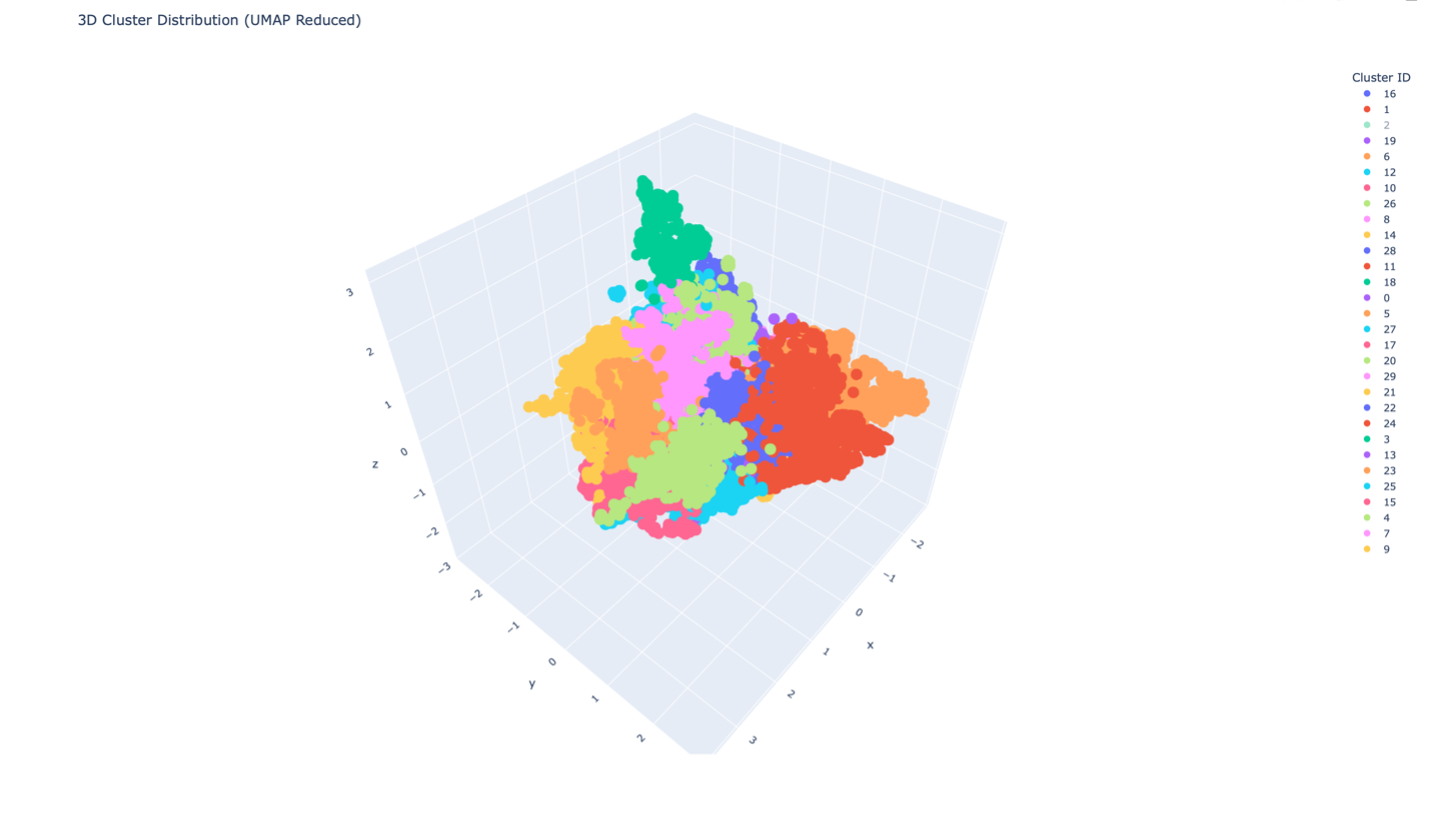}
  \caption{\textbf{Task space visualization.}
  Low-dimensional projection of prompt embeddings, colored by cluster assignment ($K=30$), illustrating the induced
  task-typing structure used throughout Sec.~\ref{sec:task_typing}.}
  \label{fig:task_space}
\end{figure}

\begin{figure}[t]
  \centering
  \begin{subfigure}[t]{0.48\linewidth}
    \centering
    \includegraphics[width=\linewidth]{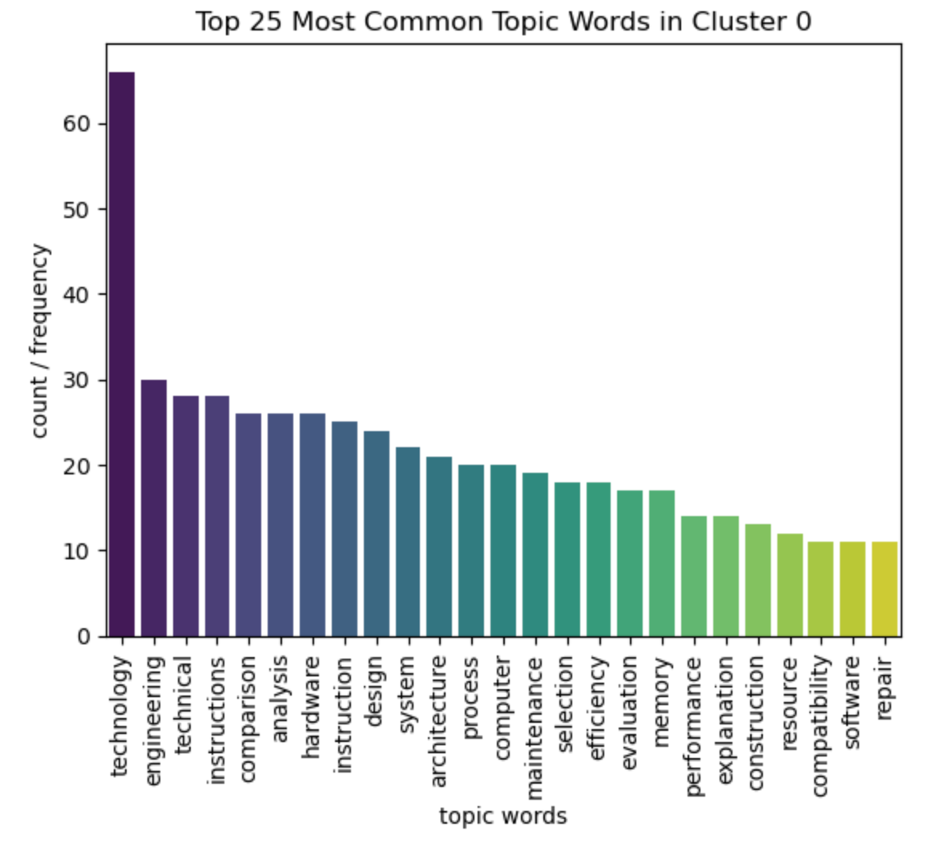}
    \caption{Example cluster A.}
    \label{fig:topic_words_a}
  \end{subfigure}\hfill
  \begin{subfigure}[t]{0.48\linewidth}
    \centering
    \includegraphics[width=\linewidth]{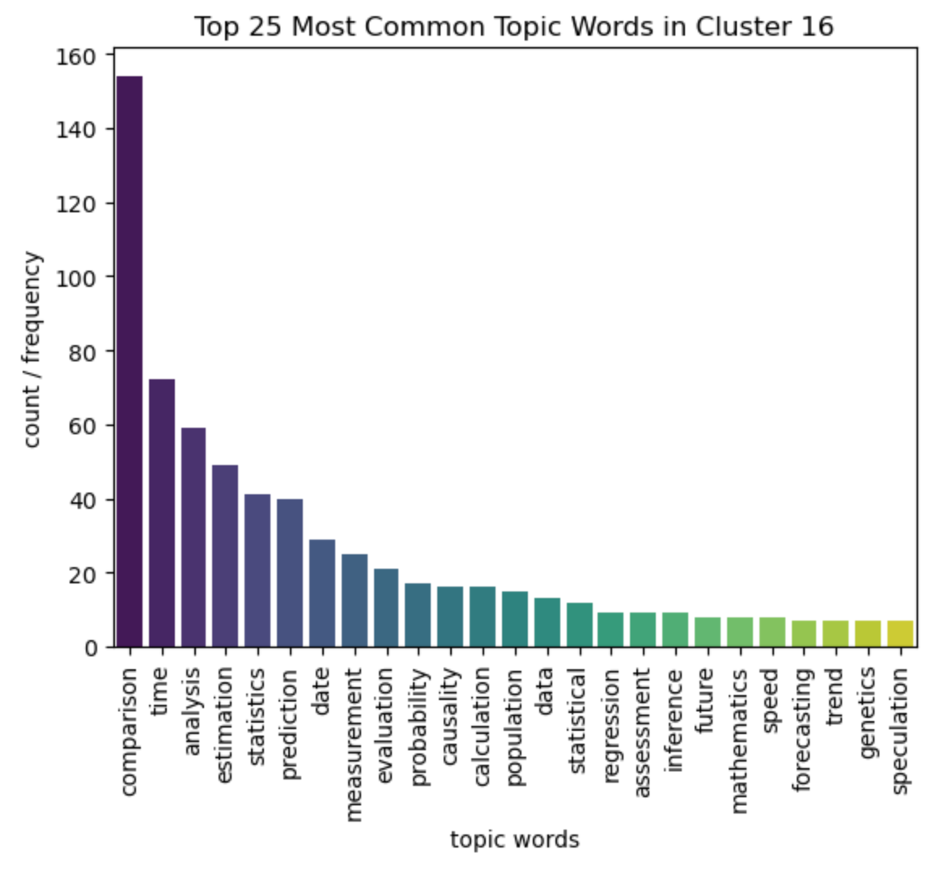}
    \caption{Example cluster B.}
    \label{fig:topic_words_b}
  \end{subfigure}
  \caption{\textbf{Interpretable task labels via cluster keywords.}
  Representative topic words for two clusters, used to assign human-readable labels and to support
  common-ground negotiation in task typing (Sec.~\ref{sec:task_typing}).}
  \label{fig:topic_words}
\end{figure}

\begin{figure}[t]
  \centering
  \includegraphics[width=\linewidth]{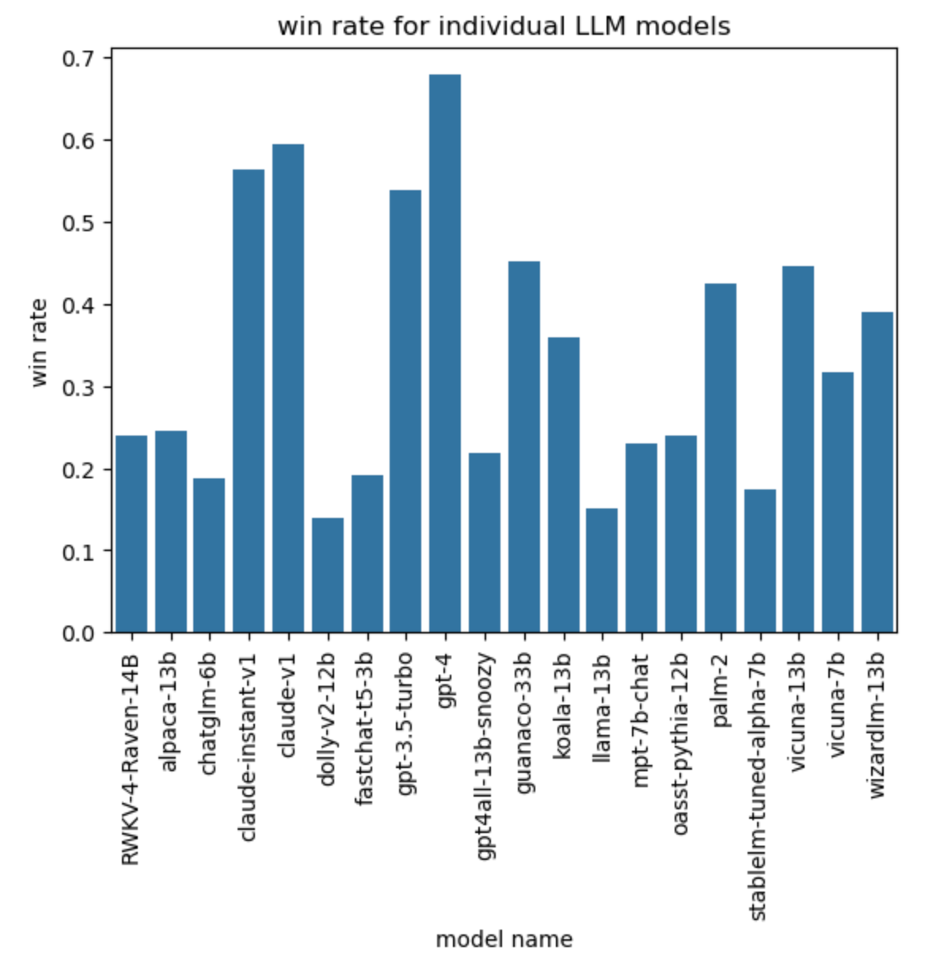}
  \caption{\textbf{Overall preference win rate across tasks.}
  Aggregate win rates across all prompts. We treat this as a global baseline; our focus is the task-conditioned
  profiles in Fig.~\ref{fig:capability_map}.}
  \label{fig:overall_winrate}
\end{figure}

\begin{figure}[t]
  \centering
  \includegraphics[width=0.95\linewidth]{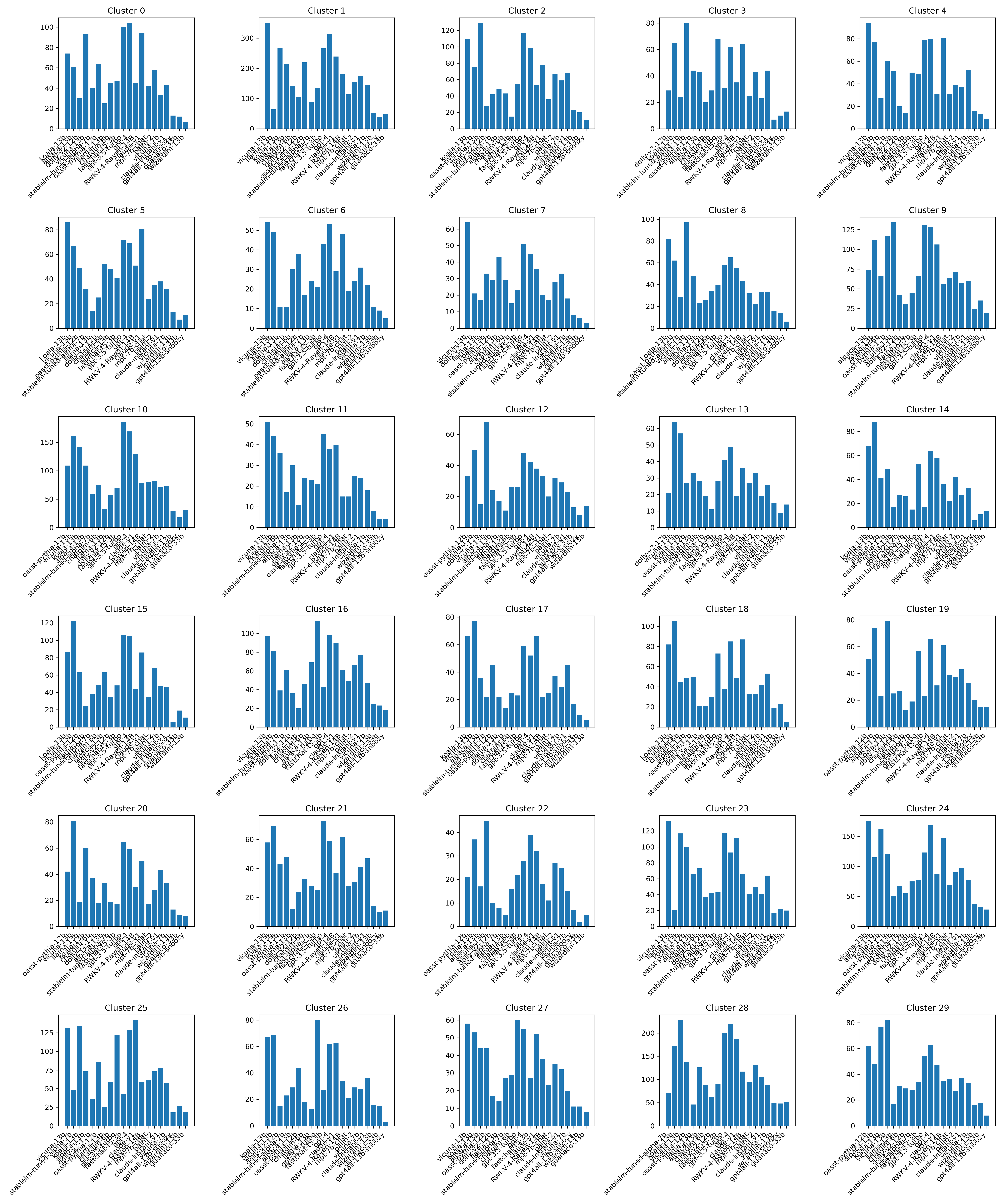}
  \caption{\textbf{Capability Profile Map (task-conditioned preference).}
  Each task cluster induces a distinct winner distribution across candidate agents, yielding a
  task-conditioned capability profile. \emph{Delegation rule:} for cluster $c$, select the top-1 agent as the primary
  collaborator; when coordination risk is high (Sec.~\ref{sec:coord_risk}), assign an auditor agent.}
  \label{fig:capability_map}
\end{figure}

\end{document}